\documentclass{elsart1p}
\usepackage{graphicx}
\usepackage{mciteplus}
\usepackage{hyperref}
\usepackage{amssymb,amsmath}
\usepackage[rflt]{floatflt}

\newcommand{\qs}{{Q_\mathrm{s}}}

\newcommand{\lqcd}{\Lambda_{\mathrm{QCD}}}
\newcommand{\as}{{\alpha_{\mathrm{s}}}}

\newcommand{\gev}{\textrm{ GeV}}

\newcommand{\ra}{R_A}

\newcommand{\npart}{{N_\textrm{part}}}

\newcommand{\ud}{\, \mathrm{d}}
\newcommand{\fig}{Fig.~}

\newcommand{\se}{Sec.~}

\begin{document}
\begin{frontmatter}

\title{Initial conditions of heavy ion collisions and small $x$}
\author[l1,l2]{T. Lappi}
\date{15 January 2009}
\address[l1]{Department of Physics
 P.O. Box 35, 40014 University of Jyv\"askyl\"a, Finland}
\address[l2]{Institut de Physique Th\'eorique,
B\^at. 774, CEA/DSM/Saclay, 91191 Gif-sur-Yvette Cedex, France}
\begin{abstract}
The Color Glass Condensate (CGC), describing the physics of the nonlinear
gluonic interactions of QCD at high energy, provides a consistent first-principles framework 
to understand the initial conditions of heavy ion collisions.
This talk reviews some aspects of the initial conditions at RHIC, and discusses implications 
for LHC heavy ion phenomenology. The CGC provides a way compute 
bulk particle production and understand recent experimental observations 
of long range rapidity correlations in terms of the classical glasma field in the 
early stages of the collision.
\end{abstract}
\end{frontmatter}
\section{Introduction}

The Relativistic Heavy Ion Collider (RHIC) at Brookhaven has been in operation since 2000
and has produced a wealth of experimental results. The combination of new theoretical
 developments with these observations has taught us a lot about different aspects of the collision
process and given much insight into what we can expect from the LHC heavy ion programme.
The focus of this talk is on the initial stages of the collision process and how it can
be understood from first principles knowledge of the high energy wavefunction of a hadron 
or a nucleus. We shall first discuss the Color Glass Condensate (CGC, 
for reviews see~\cite{Iancu:2003xm,*Weigert:2005us}) picture of the wavefunction and how
it leads to the Glasma~\cite{Lappi:2006fp} field configurations in the initial nonequilibrium 
stage of the collision. We then move on, in \se\ref{sec:bulk}, to discuss the predictions for
the total multiplicities. Section~\ref{sec:geometry} deals with more detailed aspects of the collision 
geometry and \se\ref{sec:glasma} with some more recent ideas of Glasma physics.

\begin{figure}
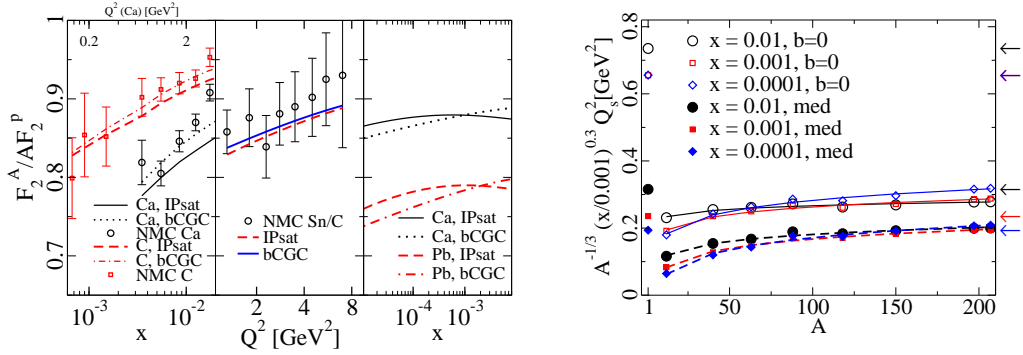

\includegraphics[width=0.5\textwidth]{shadcombowbcgcsncwide.eps} 
\hfill
\includegraphics[width=0.43\textwidth]{Qscombinationwipsat.eps} 
\caption{Left: comparison of CGC-based fits to HERA data (IPsat and bCGC models) with 
the existing eA data.
 Right: Values of the saturation scale in nuclei. For details see \cite{Kowalski:2007rw}.}
\label{fig:Qs}
\end{figure}

The central rapidity region in high energy collisions originates from the interaction 
of the ``slow'' small $x$ degrees of freedom, predominantly gluons,
 in the wavefunctions of the incoming hadrons or nuclei.
At large energies these gluons form a dense system 
that is characterized by a \emph{saturation scale} $\qs$.
The degrees of freedom with $p_T \lesssim \qs$ are fully nonlinear Yang-Mills fields with large field
strength $A_\mu \sim 1/g$ and occupation numbers $\sim 1/\as$, they can therefore be understood
as classical fields radiated from the large $x$ partons. Note that while this description is inherently 
nonperturbative, it is still based on a weak coupling argument, because the classical approximation requires 
$1/\as$ to be large and therefore $\qs \gg \lqcd$.
A ``pocket formula'' \cite{Kowalski:2007rw}
for estimating the energy and nuclear dependence of the saturation scale is 
$\qs^2 \sim A^{1/3} x^{-0.3}$: nonlinear high gluon density effects are enhanced by going to 
small $x$ and large nuclei. Ideally one would like to study the physics of the CGC at the Electron Ion 
Collider~\cite{Deshpande:2005wd}, but already based on fits to HERA data and simple nuclear geometry
we have a relatively good idea of the magnitude of $\qs$ at RHIC energies as shown in \fig\ref{fig:Qs}.
The CGC is a systematic effective theory (effective because the large $x$ part of the wavefunction is 
integrated out) formulation of these degrees of freedom, and the term glasma refers to 
the coherent, classical field configuration resulting from the collision of two such objects CGC.

\section{Bulk gluon production}
\label{sec:bulk}

\begin{figure}
\includegraphics[width=0.476\textwidth]{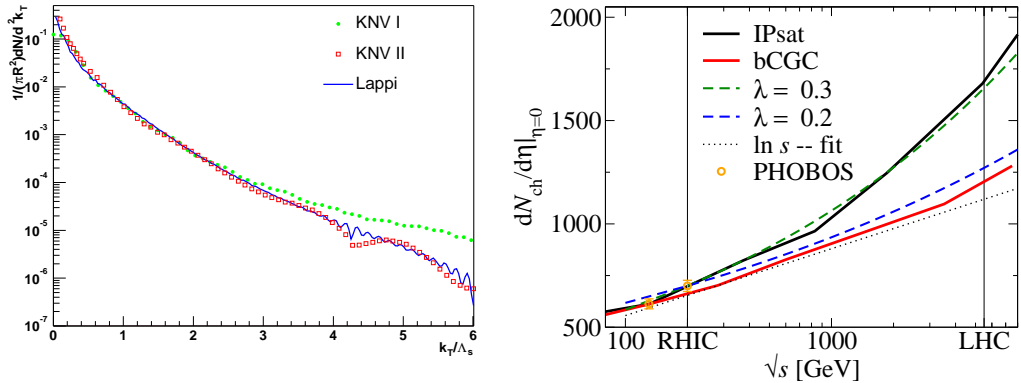}
\hfill
\includegraphics[width=0.5\textwidth]{enscan1.eps}
\caption{Left: gluon spectrum from CYM calculations~\cite{Krasnitz:2001qu,*Lappi:2003bi,*Krasnitz:2003jw}.
Right: energy dependence of the multiplicity in central collisions
based on two fits to HERA data \cite{Lappi:2008eq}.
 }
\label{fig:multi}
\end{figure}

\begin{floatingfigure}{0.4\textwidth}
\includegraphics[width=0.4\textwidth]{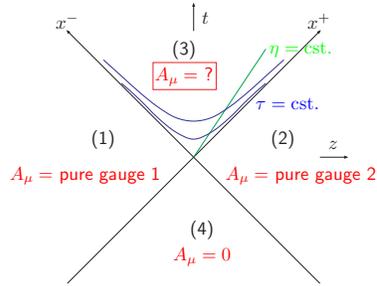}
\caption{Classical field configurations.}
\label{fig:spacet}
\end{floatingfigure}
In order to compute particle production in the Glasma one starts with the following
setup \cite{Kovner:1995ts}.
The valence-like degrees of freedom of the two nuclei are represented
by two classical color currents that are, because of their large longitudinal momenta ($p^\pm$)
well localized on the light cone (in the variables conjugate to $p^\pm$, namely $x^\mp$):
 $J^\pm \sim \delta(x^\mp)$.
These then generate the classical field that one wants to find. Working in 
light cone gauge (actually $A_\tau=0$-gauge for the two nucleus problem), the field in the region
of spacetime causally connected to only one of the nuclei is a transverse pure gauge, independently 
for each of the two nuclei. These pure gauge fields then give the initial condition on the future 
light cone ($\tau = \sqrt{2 x^+ x^-} = 0$) for the nontrivial gauge field after the collision.
The spacetime structure of these fields is illustrated in \fig\ref{fig:spacet}. The field
inside the future light cone can then be computed
either numerically~\cite{Krasnitz:2001qu,*Lappi:2003bi,*Krasnitz:2003jw} or
analytically in different approximations (see e.g.~\cite{Blaizot:2008yb} for recent work).
The obtained result is then averaged over the configurations of the sources $J^\mu$ with a distribution
$W_y[J^\mu]$ that includes the nonperturbative knowledge of the large $x$ degrees of freedom.
The resulting fields are then decomposed into Fourier modes to get the gluon spectrum, 
see \fig\ref{fig:multi}.
This is the method that we will refer to as Classical Yang-Mills (CYM) calculations.
Note that the average over configurations
 is a classical average over a probabilistic distribution. This is
guaranteed by a theorem~\cite{Gelis:2008rw,*Gelis:2008ad} ensuring
the factorization of leading logarithmic corrections to gluon production into
the quantum evolution of $W_y[J^\mu]$, analogously to the way leading logarithms of $Q^2$ 
are factorized into DGLAP-evolved parton distribution functions.

In the limit when either one or both of the color sources are dilute (the ``pp'' and ``pA'' cases), the
CYM calculation can be done analytically and reduces to a factorized form in terms of a convolution of
unintegrated parton distributions 
that can include saturation effects.
Although this approach (known as ``KLN'' after the authors of~\cite{Kharzeev:2000ph,*Kharzeev:2001gp})
is not strictly valid for the collision of two dense systems, it does have the advantage of offering
some analytical insight and making it easier to incorporate large-$x$ ingredients into the calculation.

The CYM calculations~\cite{Krasnitz:2001qu} of gluon production paint a fairly consistent picture
of gluon production at RHIC. The estimated value $\qs \approx 1.2 \gev$ from HERA 
data~\cite{Kowalski:2007rw} (corresponding to the MV model parameter 
$g^2 \mu \approx 2.1 \gev$~\cite{Lappi:2007ku}) leads to $\frac{\ud N}{\ud y} \approx 1100$ 
gluons in the initial stage. Assuming a rapid thermalization and nearly ideal hydrodynamical evolution
this is consistent with the observed $\sim 700$ charged ($\sim 1100$ total) particles produced in a unit of 
rapidity in central collisions. 

The gluon multiplicity is, across different parametrizations to a very good 
approximation proportional to $\pi \ra^2 \qs^2/\as$. Thus the predictions for LHC collisions depend mostly
on the energy dependence of $\qs$. On this front there is perhaps more uncertainty than is generally 
acknowledged, the estimates for  $\lambda= \ud \ln \qs^2 / \ud \ln 1/x$ varying between 
$\lambda= 0.29$~\cite{Golec-Biernat:1998js}
and $\lambda= 0.18$~\cite{Kowalski:2006hc} in fixed coupling fits to HERA data, with 
a running coupling solution of the BK equation giving something in between these 
values~\cite{Albacete:2007sm}. This dominates the uncertainty in predictions for the LHC multiplicity
(see \fig\ref{fig:multi}).

\begin{figure}
\includegraphics[width=0.44\textwidth]{ecc.eps}
\hfill
\includegraphics[width=0.52\textwidth]{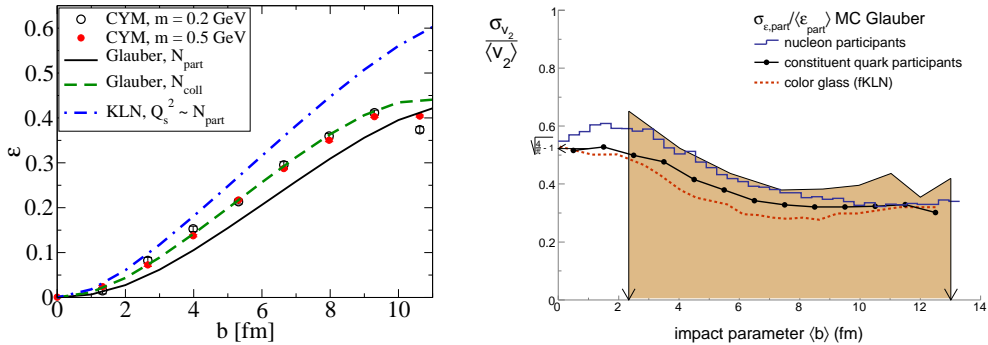}
\caption{Left: initial eccentricity from CYM~\cite{Lappi:2006xc} (``KLN'' is the result
of Ref.~\cite{Hirano:2005xf}). Right: eccentricity fluctuations from STAR compared to the
``fKLN'' model~\cite{Drescher:2007ax}.}
\label{fig:ecc}
\end{figure}

The RHIC collision energy is still too slow to clearly see any saturation effects in
the rapidity dependence of the multiplicity around $y=0$. A simple estimate for the effects
of large $x$ physics, such as momentum conservation, is to consider the typical 
$(1-x)^4$-dependence of gluon distributions at large $x$. 
Inserting $x=e^{\pm y} \langle p_\perp \rangle / \sqrt{s} $ leads to the 
estimate $\Delta y \sim \sqrt{8  \sqrt{s}/ \langle p_\perp \rangle}$ for 
the rapidity scale at which the large $x$ effects contribute to the rapidity distribution 
around $y=0$, with 
$\Delta y \sim 4$ RHIC and $\Delta y \sim 19 $ at LHC. 
The large $x$ contribution is an effect of order 1 at this scale, whereas
small $x$ evolution can be expected to give a much smaller effect~\cite{Lappi:2004sf} 
at a rapidity scale  $\Delta y \sim 1/\as \sim 3$. Only at  the LHC
the large $x$ effects will be mostly absent around midrapidity and one has a good possibility
of seeing  CGC effects in the rapidity dependence of the multiplicity.

% Without this the title is pulled very close to previous line
\vspace{0.3cm}

\section{Geometry}
\label{sec:geometry}

The basic features (such as the mostly $\npart$ scaling) of the centrality dependence of particle
multiplicities are mostly straightforward consequences of geometry and the proportionality
of the multiplicity to $\qs^2$; they are successfully reproduced by 
both KLN and CYM calculations~\cite{Kharzeev:2000ph,Lappi:2006xc,Drescher:2007ax}, 
see \fig\ref{fig:ecc}.
A striking signal of collective behavior of the matter
produced at RHIC is elliptic flow. Comparing hydrodynamical calculations with flow
is a way to address fundamental observables properties of the medium, such as viscosity,
but this comparison requires understanding of the initial conditions of the hydrodynamical
evolution, particularly the initial eccentricity for elliptic flow.
The original general consensus some years ago was that ideal hydrodynamics is in good agreement
with the experimental data, but this claim has been questioned recently after 
it was argued using a KLN-type calculation~\cite{Hirano:2005xf} that CGC
results in a larger initial eccentricity, leaving more room for viscosity in the hydrodynamical
evolution. It was subsequently pointed out in Ref.~\cite{Lappi:2006xc} that this claim of a higher 
initial eccentricity was caused partly by the unphysical nonuniversal saturation scale used
in the KLN calculation. The result of the CYM calculation~\cite{Lappi:2006xc}, later confirmed 
in a modified KLN framework~\cite{Drescher:2006ca} is that when the nonuniversality effect is
corrected for, the CGC eccentricity is indeed higher (closer to the energy density scaling with
the number of collisions) than the traditional initial conditions used in hydrodynamical
calculations (with $\npart$ scaling), but not as large as in Ref.~\cite{Hirano:2005xf}. The difference is 
illustrated in \fig\ref{fig:ecc}. A much more detailed probe of our understanding of the initial 
geometry is provided by fluctuations~\cite{Sorensen:2008zk,*Alver:2008hu} in the elliptic flow,
for some early work on the subject see Ref.~\cite{Drescher:2007ax}.

\section{Glasma physics}
\label{sec:glasma}

\begin{figure}
\includegraphics[width=0.42\textwidth,clip=true]{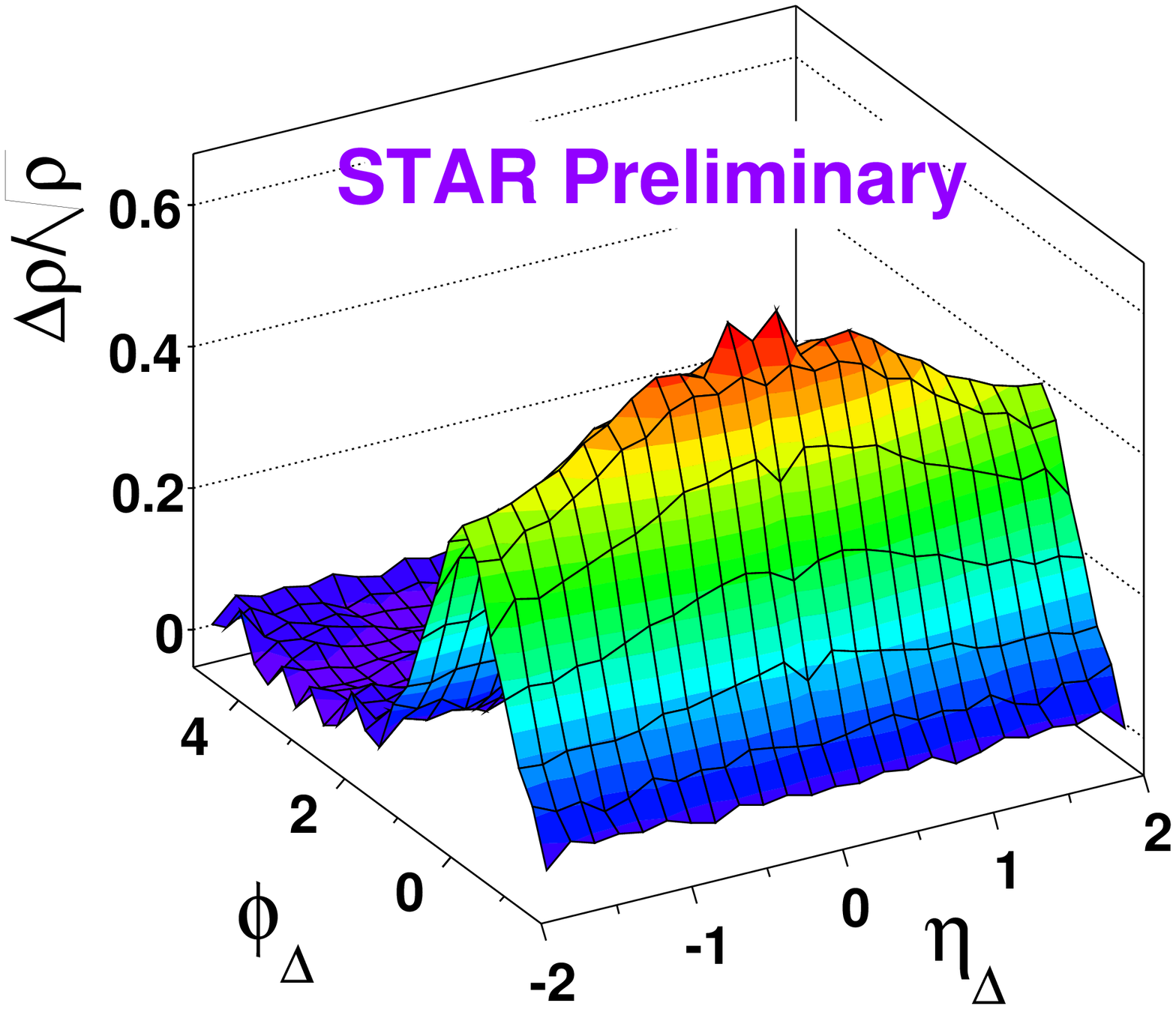}
\hfill
\includegraphics[width=0.57\textwidth,clip=true]{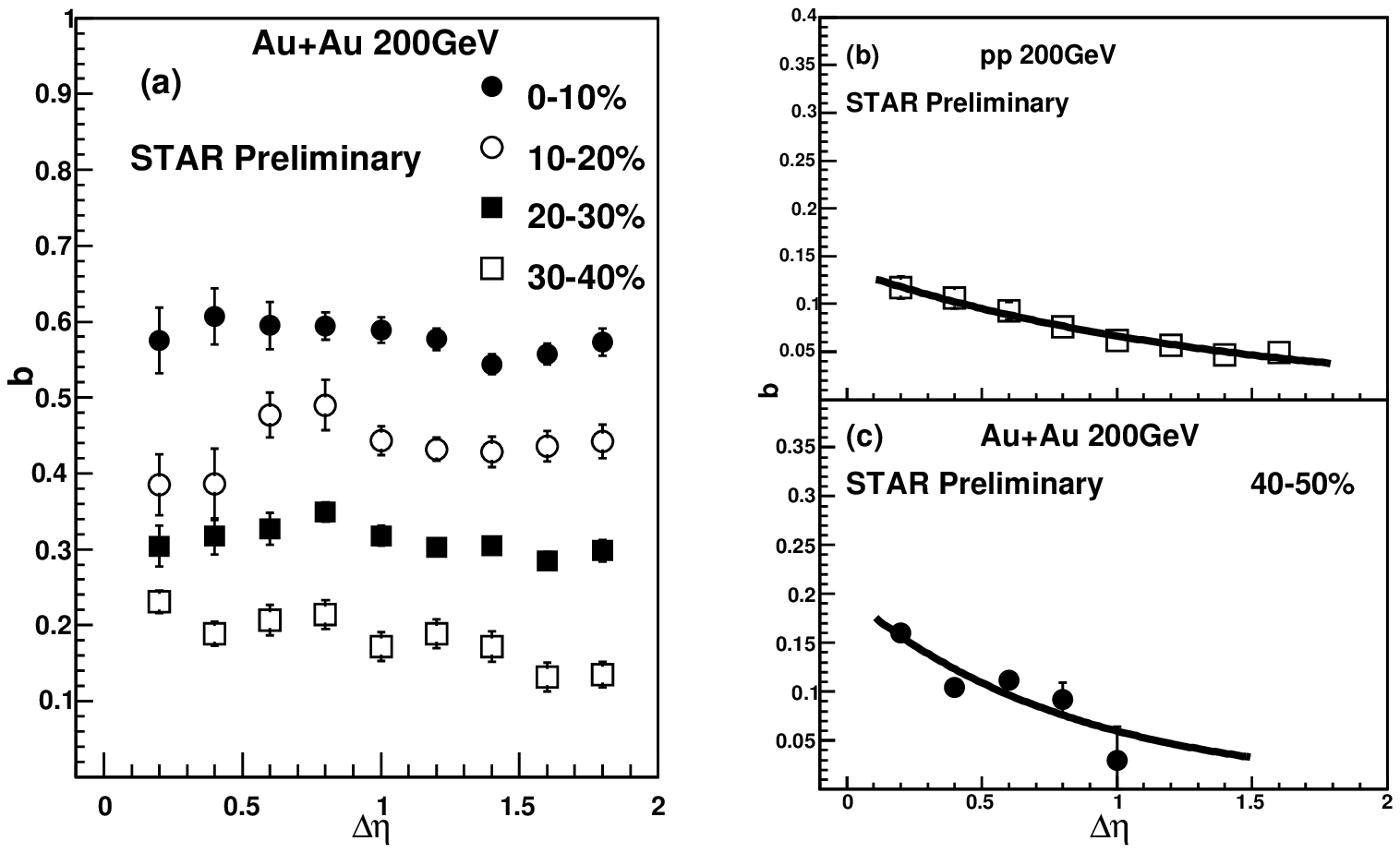}
\caption{Left: two particle correlation from STAR, showing the ``ridge'' 
structure elongated in pseudorapidity.
Right: Long range correlations in multiplicity:
$b = (\langle N_F  N_B \rangle - \langle N_F \rangle \langle N_B \rangle)
/(\langle N_F^2 \rangle - \langle N_F \rangle^2 ),$ where $F$ and $B$ are pseudorapidity
bins separated by $\Delta \eta$~\cite{Putschke:2007mi,*Daugherity:2008su,*Srivastava:2007ei}.
}
\label{fig:longrange}
\end{figure}

\begin{figure}
\includegraphics[width=0.55\textwidth]{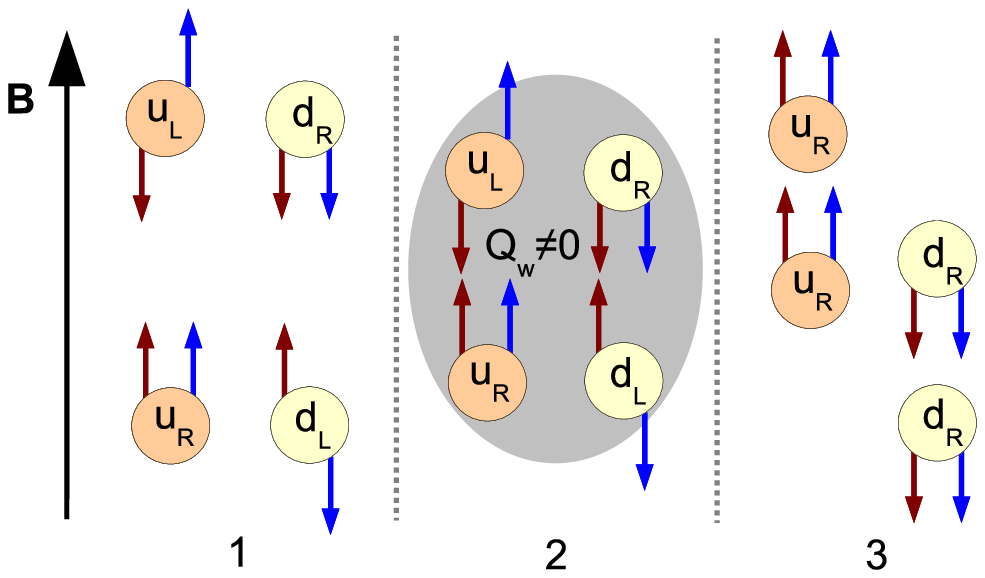}
\hfill
\includegraphics[width=0.4\textwidth]{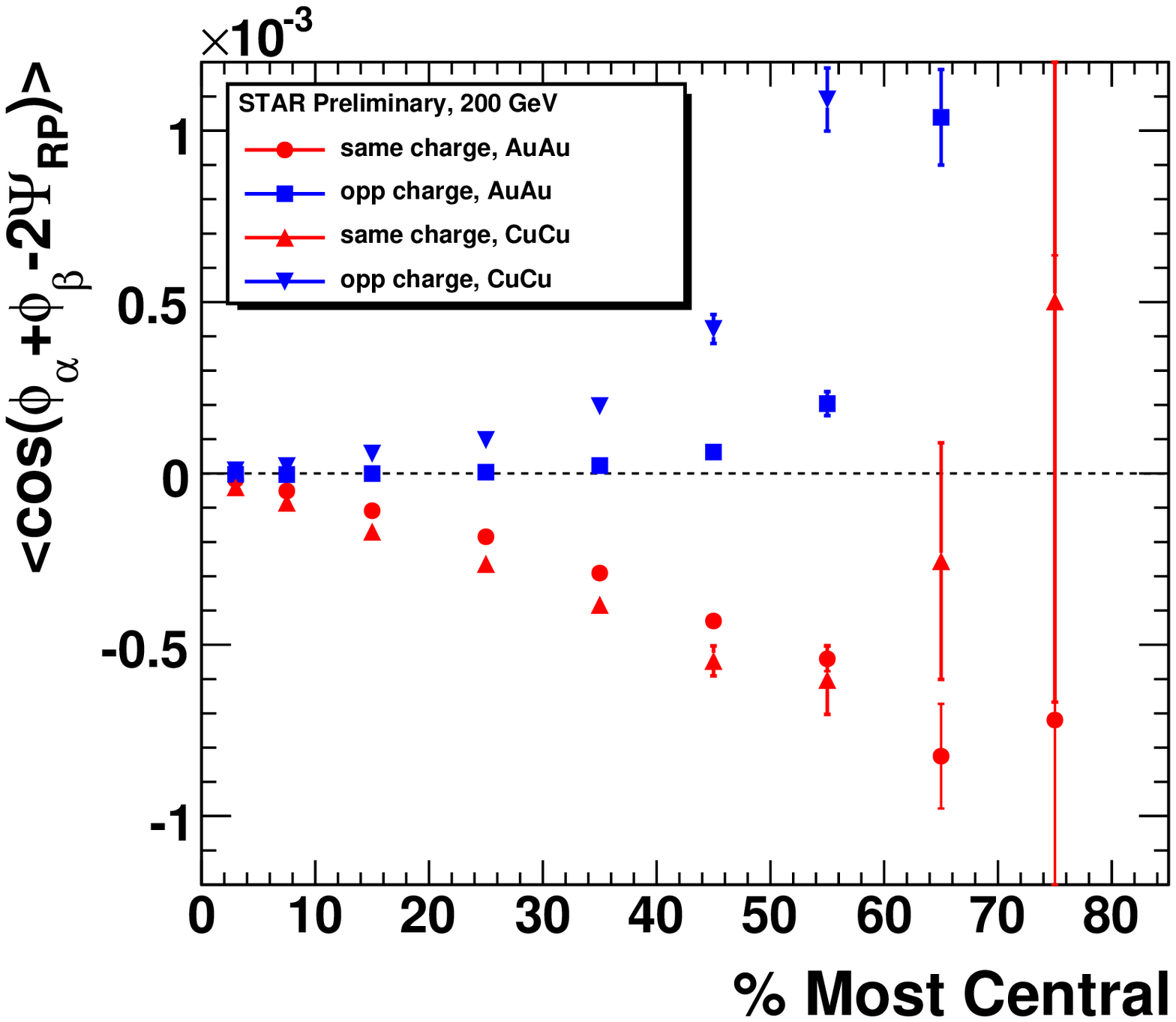}
\caption{Left: chiral magnetic effect: the blue (right) arrow is the spin
(aligned with $B$) and the red (left) the momentum, 
the evolution of the system through a configuration of nonzero Chern-Simons charge generates a 
net chirality that aligns the momenta with the spin and thus the external magnetic field, i.e. the
reaction plane. 
Right: parity violating correlation
between reaction plane and momentum~\cite{Voloshin:2008jx}. }
\label{fig:chimag}
\end{figure}

Correlations over large distances in rapidity must, by causality, originate from the earliest
times in the collision process, and are thus sensitive to the properties of the glasma phase of 
the collision. Examples of such phenomena are the 
``ridge'' and  long range correlations in
 multiplicity~\cite{Putschke:2007mi}
(see \fig\ref{fig:longrange}). The boost invariant nature of the Glasma fields
naturally leads to this kind of a correlation, and deviations from it should
be calculable from the high energy evolution governing the rapidity 
dependence~\cite{Dumitru:2008wn,*Gelis:2008sz}

Another remarkable phenomenon that is possible in  the Glasma field
configuration is the generation of a large Chern-Simons charge 
and thus parity violation~\cite{Kharzeev:2001ev,*Kharzeev:2007jp}
due to the nonperturbatively large field configuration. Through the so called
``chiral magnetic effect'' (see \fig\ref{fig:chimag}) this can manifest itself in a 
parity violating correlation between the electric dipole moment (or momentum anisotropy 
between negative and positive charges; a vector) and the reaction plane
(the positive charges of the ions generate a magnetic field perpendicular
to the reaction plane; a pseudovector). There are some preliminary indications 
in the data for such a nonzero value for such an observable~\cite{Voloshin:2008jx}.

In conclusion, RHIC data has given much quantitative insight into the small $x$ physics 
of the initial conditions, both on a quantitative level and by revealing new qualitative phenomena.
These form a good basis for LHC predictions, better than the theoretical situation was before the start 
of RHIC operations. Given the much higher energy of the LHC there are, however, also qualitatively
new phenomena in the physics of the Glasma that will only open up to experimental study at the LHC.

The author is supported by the Academy of Finland, contract 126604.

\bibliographystyle{h-physrev4mod2M}
\bibliography{spires}
\end{document}